\documentclass[aps,twocolumn,prl,preprintnumbers]{revtex4}

\input epsf

\def\be{\begin{equation}}
\def\ee{\end{equation}}
\def\bea{\begin{eqnarray}}
\def\eea{\end{eqnarray}}

\def\cmm2{{\,\rm cm^{-2}}}
\def\cm2{{\,{\rm cm}^2}}
\def\cmm3{{\,{\rm cm}^{-3}}}
\def\gcmm3{{\,{\rm g\,cm^{-3}}}}

\def\fun#1#2{\lower3.6pt\vbox{\baselineskip0pt\lineskip.9pt
  \ialign{$\mathsurround=0pt#1\hfil##\hfil$\crcr#2\crcr\sim\crcr}}}

\def\vec{\bf}

\def\eg{{e.g., }}
\def\ie{{i.e., }}

\def\p3m{P$^3$M}

\def\fun#1#2{\lower3.6pt\vbox{\baselineskip0pt\lineskip.9pt
  \ialign{$\mathsurround=0pt#1\hfil##\hfil$\crcr#2\crcr\sim\crcr}}}

\def\d{\delta}

\def\aap{Astron. \& Astrophys.}

\newcommand{\rhom}{n_{\rm M}}

\newcommand{\wde}{w_{\rm DE}}
\newcommand{\wdem}{w_{\rm tot}}

\def\pert#1{{\delta\!{#1}}}

\begin{document}
\bibliographystyle{prsty}

\preprint{UCI-TR-2006-1}
\title{Stable Models of Super-acceleration}
\author{Manoj\ Kaplinghat and Arvind\ Rajaraman}
\affiliation{Department of Physics and Astronomy\\
University of California, Irvine, California 92697, USA}
\date{\today}

\begin{abstract}
We discuss an instability in a large class of models where dark
energy is coupled to matter. In these models the mass of the
scalar field is much larger than the expansion rate of the
universe. We find models in which this  instability is absent, and
show that these models generically predict an {\em apparent}
equation of state for dark energy smaller than -1, \ie
super-acceleration. These models have no acausal behavior or
ghosts.
\end{abstract}
 \pacs{98.70.Vc} \maketitle

\section{Motivation\label{sec:motivation}}
Observations of distant Type Ia supernovae \cite{riess98,
  perlmutter99} and the cosmic microwave background \cite{spergel03}
 together strongly prefer an accelerated
expansion of the universe in the recent past.  In the standard
cosmological model this is accommodated by introducing ``dark
energy'', a component which has a significantly negative pressure
causing the expansion of the universe to accelerate.

In the standard cosmological model, dark energy is completely
decoupled from the rest of the matter in the universe except for
its gravitational effects.  It is interesting to consider more
general models in which the dark matter and dark energy have a
coupling. Such models could have new nontrivial signatures in
cosmology and structure formation.

One simple class of such models is a model in which the vacuum
energy density depends on the matter density. We shall consider a
class of these models in which the dark energy responds to changes
in the matter density on a time scale shorter than the expansion
time scale. For example, one can consider models with scalar field
dark energy coupled to matter (\eg
\cite{casas91,anderson97,amendola99,bean01,comelli03,farrar03,chimento03}),
in which the mass of the scalar field is much larger than the
expansion rate (for example, the MaVaN scenario \cite{fardon03}).

As we show below, these models generically suffer from an
instability which we label AZK-instability. The AZK-instability
was pointed out in the context of mass-varying neutrinos
(MaVaN) \cite{afshordi05}. A similar effect was identified in the
context of unified dark energy models \cite{beca05}. This
instability can also occur in models of dark energy coupled to
matter, such as the MaVaN scenario \cite{fardon03}, the Chameleon
dark energy scenario \cite{brax04} and the Cardassian expansion
scenario \cite{freese02}. Not all models in the above scenarios
are necessarily unstable (for example,
\cite{fardon06,koivisto05,takahashi06}). This will become clear
when we discuss the instability.

In this paper, we will construct a large class of models in which
this instability is avoided. We find that these models generically
predict an apparent equation of state (pressure over energy
density) $\wde$ which is less than -1 (such a phase is labeled
super-acceleration~\cite{kaplinghat03b}). That is, a model of
interacting dark energy can be incorrectly interpreted as a theory
with super-acceleration if the interactions are not taken into
account.

 For example, the coupling of dark energy to matter could be such that
 the total matter density decreases more slowly than $1/a^3$ (where
 $a$ is the scale factor of the universe). When we interpret
 observations in such a universe with a canonical matter density term
 (that decreases with expansion as $1/a^3$) and dark energy, we would
 infer an equation of state  for dark energy more negative than it
 truly is \cite{huey04,das05}. There is no physical reason why this
 inferred equation of state cannot be below -1.

This is particularly interesting because current data seem to
favor a dark energy density which is almost constant or even
increasing with time
\cite{caldwell99,schuecker02,tonry03,knop03,choudhury03,alam03,melchiorri03,majerotto04,astier06,schaefer05}.
and  exciting results can be expected in the future
\cite{weller01,frieman02,linder03,kratochvil04}. SNIa observations
currently favor a phase of super-acceleration. Future SNIa and CMB
observations have the potential to detect super-acceleration
\cite{kaplinghat03b}. No other combination has been shown to
robustly detect the signature of super-acceleration, although
combining SNIa and baryon oscillation \cite{astier06} or weak
lensing data set seem promising. Note that a measurement of just
the average equation of state \cite{saini03} is not sufficient for
this purpose \cite{maor01}. This was made explicit recently
\cite{csaki05} using a simple single scalar field model.

Scalar field models with canonical kinetic terms always produce
$\wde > -1$. Effective models with the opposite sign kinetic term
\cite{caldwell99,schulz01} imply $\wde < -1$ but are unstable
\cite{carroll03} unless more than one scalar field
\cite{feng04,guo04,hu04,wei05,urena05} or quantum effects
\cite{onemli04} are considered. Models with higher derivative terms or
scalar-tensor theories can give rise to an apparent $\wde < -1$ 
\cite{boisseau00}, but are constrained
\cite{carroll04,vikman04,abramo05}. Interpreting an alternative
gravity theory in the context of 4-d GR can also lead to
super-acceleration
\cite{mcinnes01,sahni02,pietroni02,elizalde04,nojiri05,martin06}. Some 
Cardassian models may have $\wde <-1$
\cite{wang03,koivisto04,freese05} while still satisfying the dominant
energy  condition. Another possible way to get super-acceleration with
no  instabilities is to appeal to photon-axion mixing (conversion of
photons to axions) in a universe dominated by a cosmological constant
(or quintessence) \cite{csaki04}. 

In our models, the superacceleration arises due to interactions of
dark energy and matter. Our models therefore provide
super-acceleration with none of the attendant problems that plague
most of the above models. Furthermore, the interactions are generic;
we do not need to fine-tune couplings in order to avoid theoretical
pitfalls or observational constraints. We therefore believe that
considering interactions of dark energy is the best way to generate
models of superacceleration. 

\section{AZK-instability\label{sec:instability}}

In this section we will consider a general class of models in
which the dark energy density is coupled to the non-relativistic
matter density. For an example of how this could occur, suppose
that non-relativistic matter particles are coupled to a scalar
field. Thus the local density of the matter particles can
influence the vacuum expectation value (vev) of the scalar field.
The change in the potential of the scalar then affects the dark
energy, thus coupling matter and dark energy.

In this class of models, the matter fields  will be taken to have
a matter density $\rhom$. They are coupled to a scalar field
$\chi$ (dark energy) through Yukawa like couplings. We take the
potential to be
\bea E&=&\int d^3x\ V(\chi,\rhom)\,,\\
&=&\int d^3x\ \left[ V_0(\chi)+m\rhom+\lambda  g(\chi)\rhom
\right]\,.\label{eq:defineV}\eea We will assume that \(
m_\chi^2=V''_0(\chi_0)+\lambda g''(\chi_0)\rhom\), the
mass-squared of the scalar field about its vev $\chi_0$,  is very
large so that the $\chi$ field always sits at the minimum of its
effective potential. This is the central assumption of our paper.
The mass will certainly have to be larger than the expansion rate
of the universe to be consistent with this assumption. We will
also assume that the mass is large enough to satisfy the
constraints imposed by experiments that probe the strength of a
fifth force.

In the absence of the last term, this is the potential energy of
two decoupled fluids. The first term corresponds to a cosmological
constant term (since we have assumed that the field $\chi$ is
always at the minimum). The second term is the energy density of a
dark matter fluid with density $\rhom$ and particle mass $m$.

The last term couples these two fluids, and leads to interesting
effects. In particular $\chi_0$, the value of the scalar field at
its minimum is now found by solving the equation \bea
V'_0(\chi_0)+\lambda g'(\chi_0)\rhom =0 \,,\label{eq:potmin} \eea
where $V_0'$ and $g'$ are derivatives of $V_0$ and $g$ with
respect to $\chi$. Thus $\chi_0$  is now a function of $\rhom$.

We can make the dependence of $\chi_0$ on $\rhom$ explicit in the
following way. Consider small deviations in $\rhom$. The vev of
the scalar field shifts to account for this change in $\rhom$.
Taking a further derivative, we find \bea (V''_0(\chi_0)+\lambda
g''(\chi_0)\rhom){\partial\chi_0\over\partial\rhom}+\lambda
g'(\chi_0) =0 \,.\label{eq:dchi0dn}\eea This explicitly shows how
$\chi_0$ varies as $\rhom$ varies.

In writing Eq.~\ref{eq:defineV}, we neglected the kinetic term in
comparison to the potential. This is necessary if the scalar field
is to behave as dark energy and, as we now show, consistent with
our assumption of a large mass for the scalar field. Note that
$\dot{\chi}=\dot{\rhom}\partial \chi_0/\partial \rhom$. Working
out this expression, we find that $\dot{\chi}^2/V$ for
$\chi=\chi_0$ is given by $(V_0^{\prime 2}/V
m_\chi^2)(\dot{\rhom}/m_\chi\rhom)^2$. Lets look at changes to the
scalar field potential around $\chi=\chi_0$. Unless there are
strong fine-tunings and cancellations, we will have
$V_0'\pert{\chi} < V$ and $m_\chi^2(\pert{\chi})^2/2 < V$, which
together imply that $2V_0^{\prime 2}/V m_\chi^2 < 1$. Hence the
natural expectation is that $\dot{\chi}^2/V \sim H^2/m_\chi^2$.
For large enough $m_\chi$, the kinetic term is negligible.

We now show that there is an instability in this system. We start
with a configuration where the dark matter is evenly distributed,
and the $\chi$ field is at its minimum $\chi_0$ everywhere. Now
consider small fluctuations in the matter density $\d \rhom$ which
preserve $\int_\tau d^3x\ \d \rhom=0$, i.e., the total number in
volume $\tau$. The integral is over some region $\tau$, much
smaller than the Hubble volume, over which the fluctuations are
coherent. Such a fluctuation leads to a change in the total
energy. The energy change proportional to $\d \rhom$ vanishes
because of Eq.~\ref{eq:potmin} and the condition that $\int_\tau
d^3x\ \d \rhom=0$. The energy change to next order is \bea
\pert{E}= {1\over 2}\int d^3x
(\pert{\rhom})^2\left({\partial\chi_0\over\partial\rhom}\right)\left(
m_\chi^2
 {\partial\chi_0\over\partial\rhom}
 +2\lambda
g'(\chi_0)\right)
\\
= -{1\over 2}\int d^3x
(\pert{\rhom})^2\lambda^2{\left[g'(\chi_0)\right]^2 \over m_\chi^2
}~~~\eea

Therefore the leading correction to the energy is always negative,
implying that the configuration is unstable to the growth of these
fluctuations. We dub this the AZK-instability. This instability
was first noted in the context of the MaVaN scenario
\cite{afshordi05}.

We have neglected gravity and the expansion of the universe in the
above analysis. We neglected gravity because the relevant length
scales are much smaller than the Jeans length; the instability
occurs on all scales and hence the effect is most severe on
microscopic scales. The analysis above was thus for a region
$\tau$ much smaller than that where gravity would be important. We
neglected the expansion of the universe because the relevant time
scales are much smaller than the age of the universe. In addition
our setup started with a smooth distribution of matter. For this
one must go to scales smaller than the free-streaming scale of
dark matter particles. For example, the comoving free-streaming
scale of a typical neutralino dark matter particle is of the order
of parsec. We do not study this system on larger cosmologically
relevant scales. It is, however, unlikely that the system will
still able to drive the accelerated expansion of the universe
since the generic AZK instability is intimately related to the
adiabatic sound speed of the fluid \cite{afshordi05}.

The result above assumes that the scalar field is much heavier
than the expansion rate of the universe. This constraint is easy
to satisfy and the large mass makes the model more robust to
radiative corrections (for example, see \cite{fardon06}).
Secondly, the calculation is only valid for modes which have a
wavelength much larger than $1/m_\chi$; for shorter wavelengths,
we cannot assume that the scalar field relaxes to the minimum
quickly enough.

\section{Avoiding the AZK-instability\label{sec:avoiding-instability}}

To avoid this instability, we look at more general couplings.

Consider now a model where the total energy is \bea E=\int d^3x\
\left[ V_0(\chi)+m \rhom+\lambda g(\chi)\rhom^n \right] \,,\eea
and we choose $\lambda > 0$ without loss of generality.

Again we assume that the scalar field tracks the minimum of the
potential and hence we have, \bea V'_0(\chi_0)+\lambda
g'(\chi_0)\rhom^n=0 \,,
\\
(V''_0(\chi_0)+\lambda
g''(\chi_0)\rhom^n)\left({\partial\chi_0\over\partial\rhom}\right)+\lambda
g'(\chi_0)n\rhom^{n-1}=0\,.\eea



Following our earlier calculation, we find \bea \pert{E}=
 {1\over 2}\int d^3x
\left({\pert{\rhom} \over \rhom}\right)^2 \left(-{\left[n \lambda
g'(\chi_0) \rhom^n \right]^2 \over
 m_\chi^2}\right. \nonumber \\
\left. +\lambda n(n-1)g(\chi_0)\rhom^n { \over }\right)\,,\eea

Therefore, the instability is avoided if \bea -n^2\lambda^2\rhom^n
{\left[g'(\chi_0)\right]^2 \over m_\chi^2} +n(n-1)\lambda
g(\chi_0) > 0 \,.\label{eq:condition-on-n} \eea

We note that the first term is always negative and gets large with
$\rhom$ unless $g'(\chi_0)$ decreases fast enough. Looking at the
second term we note that any value of $0 < n\leq 1$ is unstable
independent of the form of $g(\chi)$ except for the requirement
that $g(\chi_0) > 0$ which is required anyway for the potential to
be bounded from below.

A robust way to avoid the instability is to choose $n<0$, which
makes the second term positive. This is, of course, not sufficient
to guarantee the inequality in Eq.~\ref{eq:condition-on-n}. We
need the magnitude of the second term to be larger than that of
the first. This is easy to arrange. We again look at changes to
the potential as we vary $\chi$ about $\chi_0$. If the potential
is not fine-tuned to give rise to cancellations between terms in
the Taylor expansion, then $n\lambda g' \rhom^n \pert{\chi} < V$
and also $m_\chi^2(\pert{\chi})^2/2 < V$. Putting these two
expressions together yields $2n^2\lambda^2 (g')^2
\rhom^{2n}/m_\chi^2 < V \sim \lambda g \rhom^n$. Hence we see that
it is natural, if $n<0$, for the inequality in
Eq.~\ref{eq:condition-on-n} to be satisfied.



It is also possible to avoid the instability by choosing $n>1$.
However, this region of model space will be heavily constrained by
observations. In situations where the matter density gets large,
i.e., in collapsed structures, the last term in the potential
dominates. It would make the dark energy density in galaxies
large, change structure formation and clustering properties of
dark matter halos. Therefore, these kinds of models would be
tightly constrained. In order for these models to be viable,
$\lambda$ would have to be small and the model would essentially
be the same as that with two decoupled fluids.

Thus  the requirement of AZK-stability and observational
constraints naturally lead us to consider models where $n<0$. We
now  look at observational consequences of such a coupling.

\section{AZK-stability and Super-acceleration\label{sec:super-acceleration}}

The coupling term above with $n<0$ introduces a very interesting
effect: this model has super-acceleration. That is, observations
will seem to show a phase with dark energy equation of state less
than -1.

To see this, we first note that the observational quantity that is
important is the pressure. We will fit to the observations a model
with matter that scales with the expansion as $1/a^{3}$, and dark
energy with some equation of state $w_{\rm DE}$. Note that adding
or removing a component of energy density that scales as $1/a^3$
does not change the pressure of the fluid. Hence very generally
$P_{\rm tot}=P_{\rm DE}$. $P_{\rm tot}$ is defined by the equation
$\dot{V}=-3H(V+P_{tot})$ from which we find
$P_{tot}=-V_0(\chi_0)+\lambda g(\chi_0)\rhom^n(n-1)$. We set the
equation of state $\wde \equiv P_{\rm tot}/(V-m\rhom)$ and find,
\bea
  \wde =-1+{n\lambda g(\chi_0)\rhom^n\over V_0(\chi_0)
  +\lambda g(\chi)\rhom^n}\,.\eea
  Now since $n<0$, the second term is actually negative, and we have
  $w_{DE}<-1$ i.e. super-acceleration.

 We emphasize that this super-acceleration is {\it not} accompanied by
 any of the problems normally associated with theories with
 equation of state less than -1. There is no acausal behavior, and
 there are no ghosts. This is because the
super-acceleration in our model results from an interaction which
is ignored in the fitting of theory to observations. If we fit our
observations using a canonical matter density term and dark
energy, then the interaction has the effect of making the the
effective equation of state for dark energy  more negative.

\section{Sound speed\label{sec:sound-speed}}

Here we present an alternative derivation of the instability in
terms of the sound speed of the combined fluid. A negative sound
speed squared would signal instability.

On length scales much larger than $m_\chi^{-1}$, the evolution of
the system is adiabatic and hence the sound speed is
\begin{equation}\label{eq:define-c}
c_a^2 = {\dot{P}_{\rm tot} \over \dot{V}}\,.
\end{equation}

The adiabatic sound speed in this theory can then be expressed as
\begin{eqnarray}
c_a^2 & = & { \rhom \partial \wde / \partial \rhom + \wde(1+\wde)
\over
1+\wde+m\rhom/(V-m\rhom)} \,.\label{eq:c-wde}\\
&=&{\rhom \over M} \left[ {\partial^2 V(\chi_0,\rhom) \over
\partial \rhom^2} - m_\chi^2\left({\partial \chi_0 \over
\partial \rhom}\right)^2 \right]\,,\label{eq:c-V}\\
& = & { \rhom \partial \wdem / \partial \rhom + \wdem(1+\wdem)
\over
  1+\wdem}
\,,\label{eq:c-wdem}
\end{eqnarray}
where $\wdem \equiv P_{\rm tot}/V$ is the equation of state of the
total fluid.

For a universe with an accelerating expansion $\wdem < -1/3$. For
a wide class of models with $\wdem<0$ and either the $\rhom
\wdem'$ term sub-dominant or negative, we have $c_a^2 < 0$ and the
system is unstable. This is just  the AZK-instability.

Lets now look in more detail at Eq. \ref{eq:c-wde}. First,
consider the case where $\wde>-1$: the denominator is positive and
if the $\wde'$ term is sub-dominant or negative, then
AZK-instability sets in. It is clear that this instability may not
be present in models with $\wde < -1$. We also note that this
instability will likely set in well before the current epoch
because at early times $\rhom/(V-\rhom) \gg 1$. For this case
where $\wde(1+\wde) > 0$, the sign and magnitude of the
$\rhom\wde'$ term is important. In particular, the requirement
that the $\rhom\wde'$ term is sub-dominant may not be trivial to
obtain \cite{linder06}.

While the above derivation shows us how the instability arises, it
does not provide us with an intuitive understanding of what
happens to the matter. In order to better understand that we look
at the Boltzmann equation for the matter coupled to a scalar
field. The scalar field gives the matter a mass term that can vary
spatially and temporally. Following AZK \cite{afshordi05}, we
write down the Boltzmann equation for matter neglecting gravity
and hence only valid on small scales. These are the scales of
interest since we have assumed $m_\chi \gg H$. We write down the
first order perturbations to this equation and expand the
perturbations in plane wave modes. Denoting the effective mass of
the matter particle by $M(\chi)$ we find, \bea \omega
\pert{f}({\vec p},{\vec k}) &-&
(\gamma M)^{-1}{\vec p}\cdot{\vec k}\pert{f}({\vec p},{\vec k}) \\
&+& \gamma^{-1}\pert{M}({\vec k}){\vec k}\cdot\nabla_{\vec
p}f({\vec p})=0 \,.\label{eq:boltzmann} \eea We then find the
perturbation to the matter density $\pert{\rhom}({\vec k})$ using
the above equation. In the limit that matter is non-relativistic,
the resulting equation has a simple form. We find that the
variation in effective mass of the particle is given by
$\pert{M}({\vec k}) = (M / \rhom) c_s^2 \pert{\rhom}({\vec k})$
where we have defined $c_s=\omega/k$, the sound speed of matter.
The above equation is valid for perturbations $\pert{M}$ on all
scales at which our assumptions hold. As pointed out in
\cite{afshordi05}, there is no scale in the equation for $c_s^2$
because we are studying scales where it is correct to assume that
the scalar field adjusts to changes in the matter density, and
gravity is unimportant.

We now turn to the fluid description and write
$M=V(\chi_0,\rhom)/\partial \rhom$. Using Eq.~\ref{eq:dchi0dn} for
$d\chi_0/d\rhom$, one may then obtain perturbations in $M$ as
$\pert{M} = (M / \rhom) c_a^2 \pert{\rhom}$ where $c_a^2$ is given
by Eq.~\ref{eq:c-V}. In the framework of a scalar degree of
freedom coupled to matter, both descriptions must be valid and
hence we find that $c_s^2=c_a^2$. The instability may therefore be
analyzed in terms of $c_a^2$. All of our analyses in earlier sections  
go through if we work with $c_a^2$ and we conclude that models with
super-acceleration provide a generic way to avoid the AZK
instability. 

\section{Conclusions\label{sec:conclusions}}

In this paper, we have explored the possibility that dark energy
may interact with matter. Such a hypothesis is natural if the
explanation for dark energy requires extra scalar degrees of
freedom. Unfortunately, as we have shown here, these models suffer
from a generic instability when the mass of the scalar field  is
very large. We have verified that this instability is also present
in scalar-tensor theories where the scalar plays the role of dark
energy, and also in models with multiple scalar fields.

We then looked for models where this instability could be avoided,
and found a large class of such models. Most interestingly, we
found that in these models, the {\em apparent} equation of state
of the dark energy density is generically smaller than -1. This
super-acceleration is a result of the fact that we fit
observations with models that have non-interacting matter and dark
energy fluids.

There is a theoretical prejudice against models of $\wde<-1$ due
to their apparent theoretical problems. The observational data
certainly do not disfavor $\wde< -1$. Indeed a large region of the
parameter space allowed by SNIa observations corresponds to a
constant  $\wde < -1$. Here we have shown that stable models with
$\wde< -1$ may be constructed without  encountering ghosts or
acausal behavior. These models are no more fine-tuned than
quintessence models. Thus theoretical bias against $\wde <-1$
should be treated with circumspection, and not be given any weight
when interpreting observational data.


\begin{thebibliography}{10}

\bibitem{riess98}
A.~G. Riess {\it et~al.}, Astron. J. {\bf 116},  1009  (1998).

\bibitem{perlmutter99}
S. Perlmutter, M.~S. Turner, and M.~J. White, Phys. Rev. Lett. {\bf 83},  670
  (1999).

\bibitem{spergel03}
D.~N. Spergel {\it et~al.}, Astrophys. J. Suppl. {\bf 148},  175  (2003).

\bibitem{casas91}
J.~A. Casas, J. Garcia-Bellido, and M. Quiros, Class. Quant. Grav. {\bf 9},
  1371  (1992).

\bibitem{anderson97}
G.~W. Anderson and S.~M. Carroll,   (1997), arXiv:astro-ph/9711288.

\bibitem{amendola99}
L. Amendola, Phys. Rev. {\bf D62},  043511  (2000).

\bibitem{bean01}
R. Bean, Phys. Rev. {\bf D64},  123516  (2001).

\bibitem{comelli03}
D. Comelli, M. Pietroni, and A. Riotto, Phys. Lett. {\bf B571},  115  (2003).

\bibitem{farrar03}
G.~R. Farrar and P.~J.~E. Peebles, Astrophys. J. {\bf 604},  1  (2004).

\bibitem{chimento03}
L.~P. Chimento, A.~S. Jakubi, D. Pavon, and W. Zimdahl, Phys. Rev. {\bf D67},
  083513  (2003).

\bibitem{fardon03}
R. Fardon, A.~E. Nelson, and N. Weiner, JCAP {\bf 0410},  005  (2004).

\bibitem{afshordi05}
N. Afshordi, M. Zaldarriaga, and K. Kohri, Phys. Rev. {\bf D72},  065024
  (2005).

\bibitem{beca05}
L.~M.~G. Beca and P.~P. Avelino,   (2005), astro-ph/0507075.

\bibitem{brax04}
P. Brax {\it et~al.}, Phys. Rev. {\bf D70},  123518  (2004).

\bibitem{freese02}
K. Freese and M. Lewis, Phys. Lett. {\bf B540},  1  (2002).

\bibitem{fardon06}
R. {Fardon}, A.~E. {Nelson}, and N. {Weiner}, Journal of High Energy Physics
  {\bf 3},  42  (2006).

\bibitem{koivisto05}
T. Koivisto and D.~F. Mota, Phys. Rev. {\bf D73},  083502  (2006).

\bibitem{takahashi06}
R. {Takahashi} and M. {Tanimoto}, arXiv:astro-ph/0601119  (2006).

\bibitem{kaplinghat03b}
M. Kaplinghat and S. Bridle, Phys. Rev. {\bf D71},  123003  (2005).

\bibitem{huey04}
G. Huey and B.~D. Wandelt,   (2004), astro-ph/0407196.

\bibitem{das05}
S. {Das}, P. {Stefano Corasaniti}, and J. {Khoury}, arXiv:astro-ph/0510628
  (2005).

\bibitem{caldwell99}
R.~R. Caldwell, Phys. Lett. {\bf B545},  23  (2002).

\bibitem{schuecker02}
P. Schuecker {\it et~al.}, Astron. Astrophys. {\bf 402},  53  (2003).

\bibitem{tonry03}
J.~L. Tonry {\it et~al.}, Astrophys. J. {\bf 594},  1  (2003).

\bibitem{knop03}
R.~A. Knop {\it et~al.}, Astrophys. J. {\bf 598},  102  (2003).

\bibitem{choudhury03}
T.~R. Choudhury and T. Padmanabhan, Astron. Astrophys. {\bf 429},  807  (2005).

\bibitem{alam03}
U. Alam, V. Sahni, T.~D. Saini, and A.~A. Starobinsky, Mon. Not. Roy. Astron.
  Soc. {\bf 354},  275  (2004).

\bibitem{melchiorri03}
A. Melchiorri, L. Mersini-Houghton, C.~J. Odman, and M. Trodden, Phys. Rev.
  {\bf D68},  043509  (2003).

\bibitem{majerotto04}
E. Majerotto, D. Sapone, and L. Amendola, arXiv:astro-ph/0410543  (2004).

\bibitem{astier06}
P. {Astier} {\it et~al.}, \aap {\bf 447},  31  (2006).

\bibitem{schaefer05}
B.~E. {Schaefer}, Bulletin of the American Astronomical Society {\bf
  37}, 1418 (2005).

\bibitem{weller01}
J. Weller and A. Albrecht, Phys. Rev. {\bf D65},  103512  (2002).

\bibitem{frieman02}
J.~A. Frieman, D. Huterer, E.~V. Linder, and M.~S. Turner, Phys. Rev. {\bf
  D67},  083505  (2003).

\bibitem{linder03}
E.~V. Linder \& SNAP Collaboration, Bulletin of the American
Astronomical Society {\bf 37}, 1282 (2005).

\bibitem{kratochvil04}
J. Kratochvil, A. Linde, E.~V. Linder, and M. Shmakova, JCAP {\bf 0407},  001
  (2004).

\bibitem{saini03}
T.~D. Saini, T. Padmanabhan, and S. Bridle, Mon. Not. Roy. Astron. Soc. {\bf
  343},  533  (2003).

\bibitem{maor01}
I. Maor, R. Brustein, J. McMahon, and P.~J. Steinhardt, Phys. Rev. {\bf D65},
  123003  (2002).

\bibitem{csaki05}
  C.~Csaki, N.~Kaloper and J.~Terning, JCAP {\bf 0606}, 022 (2006).

\bibitem{schulz01}
A.~E. Schulz and M.~J. White, Phys. Rev. {\bf D64},  043514  (2001).

\bibitem{carroll03}
S.~M. Carroll, M. Hoffman, and M. Trodden, Phys. Rev. {\bf D68},  023509
  (2003).

\bibitem{feng04}
B. Feng, X.-L. Wang, and X.-M. Zhang, Phys. Lett. {\bf B607},  35  (2005).

\bibitem{guo04}
Z.-K. Guo, Y.-S. Piao, X.-M. Zhang, and Y.-Z. Zhang, Phys. Lett. {\bf B608},
  177  (2005).

\bibitem{hu04}
W. Hu, Phys. Rev. {\bf D71},  047301  (2005).

\bibitem{wei05}
H. Wei and R.-G. Cai, Phys. Lett. {\bf B634},  9  (2006).

\bibitem{urena05}
L.~A. Urena-Lopez, JCAP {\bf 0509},  013  (2005).

\bibitem{onemli04}
V.~K. Onemli and R.~P. Woodard, Phys. Rev. {\bf D70},  107301  (2004).

\bibitem{boisseau00}
B. Boisseau, G. Esposito-Farese, D. Polarski, and A.~A. Starobinsky, Phys. Rev.
  Lett. {\bf 85},  2236  (2000).

\bibitem{carroll04}
S.~M. Carroll, A. De~Felice, and M. Trodden, Phys. Rev. {\bf D71},  023525
  (2005).

\bibitem{abramo05}
L.~R. Abramo and N. Pinto-Neto,   (2005), astro-ph/0511562.

\bibitem{vikman04}
A. Vikman, Phys. Rev. {\bf D71},  023515  (2005).

\bibitem{mcinnes01}
B. McInnes, JHEP {\bf 08},  029  (2002).

\bibitem{sahni02}
V. Sahni and Y. Shtanov, JCAP {\bf 0311},  014  (2003).

\bibitem{pietroni02}
M. Pietroni, Phys. Rev. {\bf D67},  103523  (2003).

\bibitem{elizalde04}
E. Elizalde, S. Nojiri, and S.~D. Odintsov, Phys. Rev. {\bf D70},  043539
  (2004).

\bibitem{nojiri05}
S. Nojiri and S.~D. Odintsov, Phys. Lett. {\bf B631},  1  (2005).

\bibitem{martin06}
J. {Martin}, C. {Schimd}, and J.-P. {Uzan}, Physical Review Letters {\bf 96},
  061303  (2006).

\bibitem{wang03}
Y. Wang, K. Freese, P. Gondolo, and M. Lewis, Astrophys. J. {\bf 594},  25
  (2003).

\bibitem{koivisto04}
T. Koivisto, H. Kurki-Suonio, and F. Ravndal, Phys. Rev. {\bf D71},  064027
  (2005).

\bibitem{freese05}
K. Freese, New Astron. Rev. {\bf 49},  103  (2005).

\bibitem{csaki04}
C.~Csaki, N.~Kaloper and J.~Terning, Annals Phys.\  {\bf 317}, 410 (2005).

\bibitem{linder06}
E.~V. Linder, Phys. Rev. {\bf D73},  063010  (2006).

\end{thebibliography}

\end{document}